# High Dynamic Range 3-Moduli Set with Efficient Reverse Converter

A. Hariri, R. Rastegar, K. Navi

*Abstract*-Residue Number System (RNS) is a valuable tool for fast and parallel arithmetic. It has a wide application in digital signal processing, fault tolerant systems, etc. In this work, we introduce the 3-moduli set $\{2^n, 2^{2n}-1, 2^{2n}+1\}$ and propose its residue to binary converter using the Chinese Remainder Theorem. We present its simple hardware implementation that mainly includes one Carry Save Adder (CSA) and a Modular Adder (MA). We compare the performance and area utilization of our reverse converter to the reverse converters of the moduli sets $\{2^n-1, 2^n, 2^n+1, 2^{2n}+1\}$ and $\{2^n-1, 2^n, 2^n+1, 2^n-2^{(n+1)/2}+1, 2^n+2^{(n+1)/2}+1\}$ that have the same dynamic range and we demonstrate that our architecture is better in terms of performance and area utilization. Also, we show that our reverse converter is faster than the reverse converter of $\{2^n-1, 2^n, 2^n+1\}$ for dynamic ranges like 8-bit, 16-bit, 32-bit and 64-bit however it requires more area.

## I. INTRIDUCTION

Residue Number System (RNS) arithmetic is a valuable tool for theoretical studies of fast arithmetic [5]. With its carry-free operations, parallelism and fault tolerance, RNS has been used in computer arithmetic since 1950s. These properties have made it very useful in some applications including digital signal processing and fault tolerant systems [4]. Different moduli sets have been presented for RNS that have different properties with regards to reverse conversion (Residue to Binary or R/B), Dynamic Range (DR) and arithmetic operations. The moduli of the forms $2^n$, $2^n-1$ and $2^n+1$ are very popular according to their easy arithmetic operations. The most famous moduli set is $\{2^n-1, 2^n, 2^n+1\}$ and several methods have been proposed for its reverse conversion and the best method has been outlined in [11]. On the other hand, there are some other moduli sets that have greater dynamic ranges in comparison with this moduli set. They include; the moduli sets $\{2^n-1, 2^n, 2^n+1, 2^{n+1}-1\}$ [2] and $\{2^n-1, 2^n, 2^n+1, 2^{n+1}+1\}$ [3] that have the dynamic ranges of $4n$ and $4n+1$ bits respectively. In [1], moduli set $\{2^n-1, 2^n, 2^n+1, 2^{2n}+1\}$ has been proposed that provides the dynamic range of $2^n \times (2^{4n}-1)$. It has been shown that the reverse converter of this moduli set has superior area-time complexity in comparison with the reverse converters of [2] and [3]. In [9] the moduli set $\{2^n, 2^n-1, 2^n+1, 2^n-2^{(n+1)/2}+1, 2^n+2^{(n+1)/2}+1\}$ has been focused on which has the same dynamic range of $2^n \times (2^{4n}-1)$ and a new reverse converter has been proposed that is more efficient than the previous converters including [8] and [10]. In this paper, we introduce the moduli set $\{2^n, 2^{2n}-1, 2^{2n}+1\}$ that has the same dynamic range as [1] and [9] but the reverse conversion can be carried out faster and it requires lower hardware area in comparison with [1] and [9]. Our reverse converter is faster than the reverse converter of [11] for dynamic ranges like 8-bit, 16-bit, 32-bit and 64-bit however it utilizes more area than the reverse converter of [11].

In Section II of this paper we provide a short background for RNS and also introduce the moduli set $\{2^n, 2^{2n}-1, 2^{2n}+1\}$. In Section III, we present two lemmas and consider the reverse conversion scheme for the proposed moduli set using the presented lemmas and the CRT. In Section IV, we provide the hardware implementation of the reverse converter and in Section V we evaluate this converter and compare the results with similar works. Finally, in Section VI we present our conclusions.

## II. BACKGROUND

RNS is defined by a set $S$ of $N$ integers that are pair-wise relatively prime. That is

$$S = \{m_1, m_2, ..., m_N\}$$

where gcd $(m_i, m_j) = 1$ for $i, j = 1...N$ and $i \neq j$ and gcd means the greatest common divisor.

Every integer $X$ in $[0, M-1]$ can be uniquely represented with a $N$-tuple where,

$$M = \prod_{i=1}^{N} m_i \quad, X \rightarrow (R_1, R_2, ..., R_N)$$

and $R_i = |X|_{m_i} = (X \bmod m_i)$; for $i=1$ to $N$

The set $S$ and the number $R_i$ are called the moduli set and the residue of $X$ modulo $m_i$ respectively. The arithmetic operations can be carried out independently for each modulo, that is

$$(x_1, x_2, ..., x_N) \bullet (y_1, y_2, ..., y_N) =$$
$$(|x_1 \bullet y_1|_{m_1}, |x_2 \bullet y_2|_{m_2}, ..., |x_N \bullet y_N|_{m_N})$$

A. Hariri and K. Navi are with Shahid Beheshti University, Tehran, Iran (e-mail: {hariri, navi}@ieee.org).

R. Rastegar is with Southern Illinois University, Carbondale, IL 62901, USA (e-mail: rrastegar@ieee.org).

where ● denotes one of the arithmetic operations of addition, subtraction, and multiplication.

Here, we propose the new moduli set $\{2^n, 2^{2n}-1, 2^{2n}+1\}$ and first, we show that this set meets the requirements of an RNS moduli set.

*Theorem 1*: The set $\{2^n, 2^{2n}-1, 2^{2n}+1\}$ is a moduli set for RNS.

*Proof:* We should show that the moduli are pair-wise relatively prime for any natural number $n$. Obviously, the first modulo is relatively prime to the other moduli therefore we only show that the second and the third moduli are relatively prime. We assume that gcd $(2^{2n}-1, 2^{2n}+1) = d$ then we have
$d \mid (2^{2n}-1)$ and $d \mid (2^{2n}+1)$
therefore,
$d \mid (2^{2n}-1+2^{2n}+1)$ so $d \mid (2 \times 2^{2n})$ or we have $d \mid (2^{2n+1})$
so $d = 1$ or $d = 2^w$ $(w \geq 1)$ but we know that $d \neq 2^w$ because $2^{2n}-1$ and $2^{2n}+1$ are odd numbers so gcd $(2^{2n}-1, 2^{2n}+1) = d = 1$. □

So our proposed moduli set can be used in RNS and we can consider its reverse converter.

III. REVERSE CONVERTER

In this section, we present the reverse converter of the moduli set $\{2^n, 2^{2n}-1, 2^{2n}+1\}$ but first, we provide two lemmas which are based on the properties that have been used in calculating the reverse converters [1][4][11].

*Lemma 1:* The residue of a negative residue number $(-v)$ in modulo $(2^n-1)$ is calculated by the one's complement operation where $0 \leq v < 2^n-1$.

*Lemma 2:* The multiplication of a residue number $v$ by $2^P$ in modulo $(2^n-1)$ is carried out by $P$-bit circular left shift where $P$ is a natural number.

Now, to calculate the number $X$ from its residues, we can apply the CRT. The CRT is formulated as;

$$X = \left| \sum_{i=1}^{N} \hat{m}_i \left| \hat{m}_i^{-1} \times R_i \right|_{m_i} \right|_M \quad (1)$$

where
$M = \prod_{i=1}^{N} m_i$ ; $\hat{m}_i = \dfrac{M}{m_i}$ ; $\left| \hat{m}_i^{-1} \times \hat{m}_i \right|_{m_i} = 1$

and $R_i = |X|_{m_i}$

Assuming $m_1 = 2^n$, $m_2 = 2^{2n}-1$ and $m_3 = 2^{2n}+1$ we have

$$\hat{m}_1 = (2^{4n}-1); \; \hat{m}_2 = 2^n(2^{2n}+1); \; \hat{m}_3 = 2^n(2^{2n}-1) \quad (2)$$

*Theorem 2*: For the proposed moduli set, we have

$$\left| \hat{m}_1^{-1} \right|_{m_1} = \left| -1 \right|_{m_1} \quad (3)$$

$$\left| \hat{m}_2^{-1} \right|_{m_2} = \left| 2^{n-1} \right|_{m_2} \quad (4)$$

$$\left| \hat{m}_3^{-1} \right|_{m_3} = \left| 2^{n-1} \right|_{m_3} \quad (5)$$

*Proof*: For (3) we have:
$\left| -1 \times (2^{4n}-1) \right|_{2^n} = \left| -2^{4n}+1 \right|_{2^n} = 1$.

for (4) we have
$\left| 2^{n-1} \times 2^n(2^{2n}+1) \right|_{(2^{2n}-1)} =$
$\left| 2^{n-1} \times 2^n(2^{2n}-1) + 2^{n-1} \times 2 \times 2^n \right|_{(2^{2n}-1)} = \left| 2^{2n} \right|_{(2^{2n}-1)} = 1$.

and for (5) we write
$\left| 2^{n-1} \times 2^n(2^{2n}-1) \right|_{(2^{2n}+1)} =$
$\left| 2^{n-1} \times 2^n(2^{2n}+1) - 2^{n-1} \times 2 \times 2^n \right|_{(2^{2n}+1)} = \left| -2^{2n} \right|_{2^{2n}+1} = 1$ □

Equation (1) can be rewritten as

$$X = \left| \sum_{i=1}^{N} \hat{m}_i \left| \hat{m}_i^{-1} \times R_i \right|_{m_i} \right|_M = \quad (6)$$

$$\sum_{i=1}^{N} \hat{m}_i \left| \hat{m}_i^{-1} \right|_{m_i} \times R_i - M \times K$$

where $K$ is an integer number and depends on the value of $X$. By replacing (2)-(5) in (6) we have:

$$X = \begin{pmatrix} (2^{4n}-1) \times (-1) \times R_1 + \\ 2^n \times (2^{2n}+1) \times 2^{n-1} \times R_2 + \\ 2^n \times (2^{2n}-1) \times 2^{n-1} \times R_3 \end{pmatrix} - M \times K \quad (7)$$

By dividing the both side of (7) by $2^n$ and calculating the floor values in modulo $(2^{4n}-1)$ we have

$$\left\lfloor \dfrac{X}{2^n} \right\rfloor = \left| \begin{array}{l} \left| -2^{3n} \times R_1 \right|_{(2^{4n}-1)} + \left| (2^{2n}+1) \times 2^{n-1} \times R_2 \right|_{(2^{4n}-1)} \\ + \left| (2^{2n}-1) \times 2^{n-1} \times R_3 \right|_{(2^{4n}-1)} \end{array} \right|_{(2^{4n}-1)} \quad (8)$$

In this case the number X can be computed by

$$X = \left\lfloor \dfrac{X}{2^n} \right\rfloor \times 2^n + R_1 \quad (9)$$

Equation (8) can be written as

$$\left\lfloor \dfrac{X}{2^n} \right\rfloor = \left| \begin{array}{l} \left| -2^{3n} \times R_1 \right|_{(2^{4n}-1)} + \\ \left| (2^{3n-1}+2^{n-1}) \times R_2 \right|_{(2^{4n}-1)} \\ + \left| (2^{3n-1}-2^{n-1}) \times R_3 \right|_{(2^{4n}-1)} \end{array} \right|_{(2^{4n}-1)} \quad (10)$$

or

$$\left\lfloor \dfrac{X}{2^n} \right\rfloor = \left| S_1 + S_2 + S_3 \right|_{(2^{4n}-1)} \quad (11)$$

where

$$S_1 = \left| -2^{3n} \times R_1 \right|_{(2^{4n}-1)} \quad (12)$$

$$S_2 = \left|(2^{3n-1} + 2^{n-1}) \times R_2\right|_{(2^{4n}-1)} \tag{13}$$

$$S_3 = \left|(2^{3n-1} - 2^{n-1}) \times R_3\right|_{(2^{4n}-1)} \tag{14}$$

Now, we consider (12)-(14) and simplify them for implementation in a VLSI system. It is necessary to note that $r_{i,j}$ means the $j$-th bit of $R_i$.

*Evaluation of $S_1$:*

The residue $R_1$ can be represented in $4n$ bits as follows;

$$R_1 = \underbrace{00\cdots00}_{3n \text{ Bits}} r_{1,(n-1)} \cdots r_{1,1} r_{1,0} \tag{15}$$

by applying Lemma 2 in modulo ($2^{4n}-1$) we have

$$\left|2^{3n} \times R_1\right|_{(2^{4n}-1)} = r_{1,(n-1)} r_{1,(n-2)} \cdots r_{1,1} r_{1,0} \underbrace{00\cdots00}_{3n \text{ Bits}} \tag{16}$$

and finally by applying Lemma 1 we have

$$S_1 = \left|-2^{3n} \times R\right|_{(2^{4n}-1)} = \overline{r}_{1,(n-1)} \cdots \overline{r}_{1,1} \overline{r}_{1,0} \underbrace{11\cdots11}_{3n \text{ Bits}} \tag{17}$$

where $\overline{r}$ means the complement of $r$.

*Evaluation of $S_2$:*

The residue $R_2$ can be represented in $4n$ bits as follows;

$$R_2 = \underbrace{00\cdots00}_{2n \text{ Bits}} r_{2,2n-1} \cdots r_{2,1} r_{2,0} \tag{18}$$

we evaluate the two parts of $S_2$ separately using Lemma 2

$$\left|2^{3n-1} \times R_2\right|_{(2^{4n}-1)} = \underbrace{r_{2,n} \cdots r_{2,1} r_{2,0}}_{n+1 \text{ Bits}} \underbrace{00\cdots00}_{2n \text{ Bits}} \underbrace{r_{2,(2n-1)} \cdots r_{2,(n+1)}}_{n-1 \text{ Bits}} \tag{19}$$

$$\left|2^{n-1} \times R_2\right|_{(2^{4n}-1)} = \underbrace{00\cdots00}_{n+1 \text{ Bits}} \underbrace{r_{2,(2n-1)} \cdots r_{2,1} r_{2,0}}_{2n \text{ Bits}} \underbrace{00\cdots00}_{n-1 \text{ Bits}} \tag{20}$$

by adding (19) and (20) we have the final value of $S_2$ as

$$S_2 = \left|2^{3n-1} \times R_2 + 2^{n-1} \times R_2\right|_{(2^{4n}-1)} = \tag{21}$$

$$\underbrace{r_{2,n} \cdots r_{2,1} r_{2,0}}_{n+1 \text{ Bits}} \underbrace{r_{2,(2n-1)} \cdots r_{2,1} r_{2,0}}_{2n \text{ Bits}} \underbrace{r_{2,(2n-1)} \cdots r_{2,(n+1)}}_{n-1 \text{ Bits}}$$

that is a $4n$-bit residue number.

*Evaluation of $S_3$:*

The residue $R_3$ can be represented in $4n$ bits as follows;

$$R_3 = \underbrace{00\cdots\cdots00}_{2n-1 \text{ Bits}} r_{3,2n} \cdots\cdots r_{3,1} r_{3,0} \tag{22}$$

for the two parts of $S_3$ we use Lemma 2 and we write

$$\left|2^{3n-1} \times R_3\right|_{(2^{4n}-1)} = \underbrace{r_{3,n} \cdots r_{3,1} r_{3,0}}_{n+1 \text{ Bits}} \underbrace{0\cdots0}_{2n-1 \text{ Bits}} \underbrace{r_{3,2n} \cdots r_{3,(n+1)}}_{n \text{ Bits}} \tag{23}$$

$$\left|2^{n-1} \times R_3\right|_{(2^{4n}-1)} = \underbrace{00\cdots00}_{n \text{ Bits}} \underbrace{r_{3,(2n)} \cdots r_{3,1} r_{3,0}}_{2n+1 \text{ Bits}} \underbrace{00\cdots00}_{n-1 \text{ Bits}} \tag{24}$$

for (24) we apply Lemma 1 and we have

$$\left|-2^{n-1} \times R_3\right|_{(2^{4n}-1)} = \underbrace{11\cdots11}_{n \text{ Bits}} \underbrace{\overline{r}_{3,(2n)} \cdots \overline{r}_{3,1} \overline{r}_{3,0}}_{2n+1 \text{ Bits}} \underbrace{11\cdots11}_{n-1 \text{ Bits}} \tag{25}$$

therefore,

$$S_{3,1} = \underbrace{r_{3,n} r_{3,(n-1)} \cdots r_{3,1} r_{3,0}}_{n+1 \text{ Bits}} \underbrace{00\cdots00}_{2n-1 \text{ Bits}} \underbrace{r_{3,2n} r_{3,(2n-1)} \cdots r_{3,(n+1)}}_{n \text{ Bits}} \tag{26}$$

$$S_{3,2} = \underbrace{11\cdots\cdots11}_{n \text{ Bits}} \underbrace{\overline{r}_{3,(2n)} \overline{r}_{3,(2n-1)} \cdots\cdots \overline{r}_{3,1} \overline{r}_{3,0}}_{2n+1 \text{ Bits}} \underbrace{11\cdots\cdots11}_{n-1 \text{ Bits}} \tag{27}$$

so, $S_3$ includes two $4n$-bit numbers that are $S_{3,1}$ and $S_{3,2}$.

## IV. HARDWARE IMPLEMENTAION

To implement the reverse converter, four $4n$-bit numbers should be summed up in modulo ($2^{4n}-1$). This requires a 2-level Carry Save Adder (CSA) tree that includes two $4n$-bit CSAs. Nevertheless by considering (17) and (27), it is clear that the $3n$ rightmost bits of $S_1$ and also the $n$ leftmost bits of $S_{3,2}$ are ones. So, we replace the $3n$ rightmost bits of $S_{3,2}$ with the same bits of $S_1$. Based on this manipulation, the new numbers have been shown in (28) and (29). Consequently, now $S_{3,2}$ contains $4n$ ones and we know that it is equivalent to zero in modulo ($2^{4n}-1$). Now, we have 3 numbers and therefore, the required 2-level CSA can be replaced by only one CSA.

$$S_1' = \overline{r}_{1,(n-1)} \cdots \overline{r}_{1,0} \underbrace{\overline{r}_{3,2n} \cdots \overline{r}_{3,0}}_{2n+1 \text{Bits}} \underbrace{11\cdots11}_{n-1 \text{Bits}} \tag{28}$$

$$S_{3,2} = \underbrace{11\cdots11}_{n \text{ Bits}} \underbrace{11\cdots11}_{3n \text{ Bits}} \tag{29}$$

Fig. 1 shows the hardware architecture of the reverse converter. The Operand Preparation (OP) component includes some wires and inverters and prepares the $4n$-bit numbers for the Multi Operand Modular Adder (MOMA). The CSA tree includes only one $4n$-bit CSA with End-Around Carry (EAC) [6]. The last component in MOMA is a Modular Adder (MA) and can be implemented using the methods of [6], [7] or [15]. The output of this adder is equivalent to $\left\lfloor \frac{X}{2^n} \right\rfloor$ and consequently, $X$ can be computed by using (9).

# High Dynamic Range 3-Moduli Set with Efficient Residue to Binary Converter

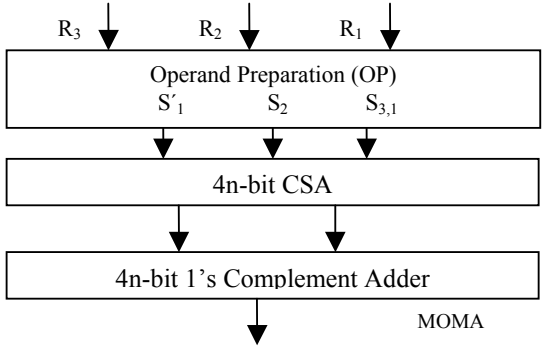

Fig 1. Hardware architecture of the proposed reverse converter

## V. EALUATION AND COMPARISON

Moduli sets of [1] and [9] provide the same dynamic range as our moduli set. So, in this section we compare two properties of our moduli set to the moduli sets of [1] and [9]; 1) Time and area complexities of the reverse conversion and 2) Time complexity of the arithmetic operations in their moduli. Finally, we compare our reverse converter to the reverse converter of a 3 moduli set proposed in [11]. Now, we compute the hardware utilization of our reverse converter in terms of adders and basic gates. As outlined in the previous section, we should sum up three $4n$-bit numbers $S'_1$, $S_2$ and $S_{3,1}$. For this purpose, one CSA which includes $4n$ Full Adders (FAs) is sufficient. But by considering the operands, it is clear that some of these FAs could be simplified further. For the ($n$-1) rightmost bits, we need ($n$-1) pairs of XNOR/OR gates instead of ($n$-1) FAs, since one of the inputs of each FA is 1. Similarly, for the middle ($2n$-1) bits, we replace the ($2n$-1) FAs with ($2n$-1) pairs of XOR/AND gates, since one of the inputs of each FA is 0. For the rest of the bits, we use ($n+2$) FAs. Besides this MOMA, the operand preparation includes some wires and inverters. Ignoring the wires, it includes ($3n+1$) inverters. The total amount of the used hardware is shown in Table I.

TABLE I
HARDWARE UTILIZATION OF THE REVERSE CONVERTER

| R/B Converter | Our work | [1] | [9] | [11] |
|---|---|---|---|---|
| DR | $2^n \times (2^{4n}-1)$ | $2^n \times (2^{4n}-1)$ | $2^n \times (2^{4n}-1)$ | $2^m \times (2^{2m}-1)$ |
| Inverters (OP) | $3n+1$ | $5n+3$ | $4n$ | $2m+1$ |
| FAs | $n+2$ | $7n+6$ | $15n$ | $2m$ |
| XOR/AND Pairs | $2n-1$ | $2n-1$ | $\approx 7n$ | - |
| XNOR/OR Pairs | $n-1$ | $4n$ | $\approx 2n$ | - |
| Other | - | $2n-3$ inverter | - | XOR+HA |
| MUX | - | - | One 4×1 | Two 2×1 |
| MA | $4n$-bit | $4n$-bit | $4n$-bit | $2m$-bit |

It is clear from Table I that our proposed reverse converter requires very low hardware area in comparison with the reverse converter of [1] and also our reverse converter is superior to the reverse converter of [9] which is the most efficient converter for the moduli set $\{2^n, 2^n-1, 2^n+1, 2^n-2^{(n+1)/2}+1, 2^n+2^{(n+1)/2}+1\}$. In [9], one 4×1 multiplexer is required for generating one of the $4n$-bit operands of the CSA tree. So this operand can have four possible values and they would only contain fixed ones and zeros. To consider its associated CSA, we have assumed that the number of ones is approximately equal to the number of zeros and this assumption does not affect the comparison. The total delay of our reverse converter is the sum of the delays of three components: the operand preparation, CSA and MA. The delay of operand preparation is equal to the delay of a NOT gate. For the CSA, the delay is the delay of an FA. For the MA, different methods can be applied that have different delays [6][7][15]. Here we have used the modular adder of [15]. Adopting the unit gate delay [11][13]15], we assume $t_{inv}=t_{and}=1$, $t_{mux}=2$, $t_{FA}=2$, $t_{xor}=2$ and consequently using the method of [15], $t_{MA(n)}=2\log_2(n)+3$. Table II shows the delays of the reverse converters. It can be concluded form Table II that we have eliminated the delay of two FAs in comparison with [1] and the delay of three FAs in comparison with [9]. In addition to this delay improvement, we have utilized much lower hardware than [1] and [9].

TABLE II
DELAYS OF THE REVERSE CONVERTERS

| R/B | Delay | Unit Gate Delay |
|---|---|---|
| [1] | $t_{CLA(4n)}+t_{NOT}+3t_{FA}$ | $2\lceil\log_2(n)\rceil + 7 + 7$ |
| [9] | $t_{CLA(4n)}+t_{NOT}+4t_{FA}$ | $2\lceil\log_2(n)\rceil + 7 + 9$ |
| [11] | $t_{CLA(2m)}+t_{NOT}+t_{MUX}+t_{FA}$ | $2\lceil\log_2(n)\rceil + 7 + 5$; if $\log_2(2n) = \log_2(m)$ (1) <br> $2\lceil\log_2(n)\rceil + 5 + 5$; if $\log_2(2n) = \log_2(m)+1$ (2) |
| Ours | $t_{CLA(4n)}+t_{NOT}+t_{FA}$ | $2\lceil\log_2(n)\rceil + 7 + 3$ |

So far, we have shown that our converter has better area and time complexities than those of [1] and [9], but we have left one question unanswered. For an equal dynamic range, is a 4 or 5-moduli set always faster than a 3-moduli set? It is the magnitude of the largest modulo that dictates the speed of arithmetic operations; however, speed and cost do not just depend on the width of the residues but also depend on the moduli chosen [5]. Consequently, for the moduli set of [1], modulo $2^{2n}+1$ determines the overall speed of the RNS. The same is true for our proposed moduli set. Therefore our moduli set and the moduli set of [1], are both restricted to the time performance of modulo $2^{2n}+1$. The moduli set of [9] includes two moduli of $(2^n-2^{(n+1)/2}+1)$ and $(2^n+2^{(n+1)/2}+1)$. Here, we compute the delay of addition in modulo $(2^n+2^{(n+1)/2}+1)$ by using the method of [11] and we compare it to delay of addition in modulo $(2^{2n}+1)$ that has been computed by using the method of [13]. Table III shows that addition in modulo $(2^{2n}+1)$ is much faster than addition in modulo $(2^n+2^{(n+1)/2}+1)$. So, we can conclude that although [9] has five moduli, it is not faster than our proposed moduli set. Therefore our moduli set

High Dynamic Range 3-Moduli Set with Efficient Residue to Binary Converter

outperforms both moduli sets of [1] and [9].

TABLE III
DELAY OF ADDITION IN TWO MODULI

| Addition in modulo ($2^n+2^{(n+1)/2}+1$) | Addition in modulo ($2^{2n}+1$) |
|---|---|
| ≈$4\times\log_2(n)+7$ | $2\times\log_2(2n)+6=2\times\log_2(n)+8$ |

In addition to comparing [1] and [9], we would like to compare our reverse converter to the reverse converters of 3-moduli sets. In [14], it has been shown that moduli set $\{2^n-1, 2^n, 2^n+1\}$ has the fastest and the most area efficient reverse converter among the other 3-moduli sets for the dynamic ranges of 8-bit, 16-bit, 32-bit and 64-bit. So, we compare our reverse converter to the reverse converter of [11] which is the most efficient reverse converter for $\{2^n-1, 2^n, 2^n+1\}$. For the sake of a fair comparison, we consider the moduli set $\{2^m-1, 2^m, 2^m+1\}$ where $m$ is chosen in a way that provides similar dynamic ranges to our moduli set and more or less $m$ can be the floor or ceiling value of $5n/3$. By using this approximation, the hardware utilization of the reverse converter of [11] has been derived and included in Table I. In Table II, we have compared our reverse converter to the reverse converter of [11] considering two cases. In case (1) our reverse converter is faster than the reverse converter of [11] and it is worthwhile to mention that for example, for $n$ in [1, 50], this case covers 73% of dynamic ranges including 8-bit, 16-bit, 32-bit and 64-bit. In case (2) which covers 26% of dynamic ranges, our reverse converter and the reverse converter of [11] have the same delay but [11] requires less hardware area. Table IV shows the area and delay comparison of the proposed reverse converter and that of the [11] using the unit-gate model where the hardware area utilization of the gates are $A_{NOT}=A_{AND}=A_{OR}=1$ and $A_{XOR}=2$. The hardware area utilization of the modular adder has been computed using the adder of [15].

TABLE IV
COMPARISON OF REVERSE CONVERSION IN TWO 3-MODULI SETS

| DR | $n$ | $m$ | $A_{ours}$ | $A_{[11]}$ | Extra Area% | $t_{ours}$ | $t_{[11]}$ | Speed-up % |
|---|---|---|---|---|---|---|---|---|
| 8-bit | 2 | 3 | 151 | 136 | 11.02 | 12 | 14 | 14.2 |
| 16-bit | 4 | 6 | 341 | 298 | 14.42 | 14 | 16 | 12.5 |
| 32-bit | 7 | 11 | 674 | 604 | 11.58 | 16 | 18 | 11.1 |
| 64-bit | 13 | 22 | 1400 | 1330 | 5.26 | 18 | 20 | 10 |

It can be concluded that the comparison of our work and [11] is purely dictated by the chosen dynamic range. However, for the discussed dynamic ranges, our reverse converter is faster than the reverse converter of [11] while [11] requires less area.

## VI. CONCLUSION

In this paper we proposed the moduli set $\{2^n, 2^{2n}-1, 2^{2n}+1\}$ and its reverse converter. This moduli set provides the dynamic range of $2^n\times(2^{4n}-1)$ and the implementation results have shown that its reverse converter has better area and time complexities in comparison with the moduli sets with the same dynamic ranges. We also showed that for majority of the similar dynamic ranges, our reverse converter is faster than the reverse converter of $\{2^n-1, 2^n, 2^n+1\}$ but the reverse converter of $\{2^n-1, 2^n, 2^n+1\}$ has less area.

ACKNOWLEDGMENT

The authors wish to acknowledge the valuable help of Dr. T. Vergos with the modular adders.